\documentclass{ws-procs10x7}

\newcommand{\ipb}{\ensuremath{\mathrm{pb^{-1}}}}

\newcommand{\TeV}{\ensuremath{\mathrm{Te\kern -0.1em V}}}
\newcommand{\TeVc}{\ensuremath{\mathrm{Te\kern -0.1em V\!/}c}}
\newcommand{\TeVcc}{\ensuremath{\mathrm{Te\kern -0.1em V\!/}c^2}}
\newcommand{\GeV}{\ensuremath{\mathrm{Ge\kern -0.1em V}}}
\newcommand{\GeVc}{\ensuremath{\mathrm{Ge\kern -0.1em V\!/}c}}
\newcommand{\GeVcc}{\ensuremath{\mathrm{Ge\kern -0.1em V\!/}c^2}}
\newcommand{\MeV}{\ensuremath{\mathrm{Me\kern -0.1em V}}}
\newcommand{\MeVc}{\ensuremath{\mathrm{Me\kern -0.1em V\!/}c}}
\newcommand{\MeVcc}{\ensuremath{\mathrm{Me\kern -0.1em V\!/}c^2}}

\newcommand{\cdfii}{CDF\,II~}

\newcommand{\ase}[2]{\ensuremath{^{~+ #1}_{~- #2}}}

\newcommand{\myto}{\kern -0.3em\to\kern -0.2em}
\newcommand{\Lxy}{L_{\rm xy}}
\newcommand{\BR}{{\mathrm {BR}}}

\newcommand{\phikpm}{\ensuremath{B^{\pm} \myto \phi K^{\pm}}}

\newcommand{\phik}{\ensuremath{B^+ \myto \phi K^+}}
\newcommand{\psik}{\ensuremath{{B^+ \myto J/\psi K^+\ }}}
\newcommand{\phiphi}{\ensuremath{B_s \myto \phi \phi}}
\newcommand{\psiphi}{\ensuremath{B_s \myto J/\psi \phi}}
\newcommand{\psikst}{\ensuremath{B_d \myto J/\psi K^{\ast 0}\ }}
\newcommand{\phikst}{\ensuremath{B_d \myto \phi K^{\ast 0}\ }}
\newcommand{\kstarkstarb}{\ensuremath{B_s \myto K^{\ast 0}\bar{K}^{\ast 0}\ }}
\newcommand{\kstar}{\ensuremath{K^{\ast 0}\ }}

\newcommand{\mpmm}{\ensuremath{\mu^+\mu^-}\ }
\newcommand{\KKpm}{\ensuremath{K^+K^-}\ }

\newcommand{\ACP}{\ensuremath{A_{CP}}}
\newcommand{\DGs}{\ensuremath{\Delta \Gamma_s}}

\newcommand{\BVV}{\ensuremath{B\rightarrow VV\ }}
\newcommand{\BsVV}{\ensuremath{B_s\rightarrow VV\ }}

\begin{document}

\title{ 
First Evidence for \phiphi\ and penguin $B$ decays at CDF
}

\author{M. Rescigno {for the CDF collaboration}}

\address{INFN, Sezione di Roma 1, P.le Aldo Moro 5, 00185 Roma, Italy\\
E-mail: marco.rescigno@roma1.infn.it}

\twocolumn[\maketitle\abstract{
We present the first evidence of the decay mode \phiphi\ and a measurement 
of partial width and direct CP asymmetry for the \phik\ decay
using $180\,{\rm pb}^{-1}$ of data collected by the \cdfii experiment 
at the Fermilab Tevatron collider. 
We measure: 
$\BR (\phiphi) = (1.4 \pm 0.6(stat.) \pm 0.2(syst.) \pm 0.5(\BR) ) \cdot 10^{-5}$,
where the last error is due to the uncertainty on 
the \psiphi\ branching ratio used as normalization,
$\BR (\phik) = (7.2 \pm 1.3(stat.) \pm 0.7(syst.) ) \cdot 10^{-6}$
and $\ACP(\phik) = -0.07 \pm 0.17 (stat.) \ase{0.06}{0.05} (syst.)$.
We also briefly discuss prospects for studying other charmless 
\BVV decays at CDF}]

\begin{figure}
\begin{center}
\includegraphics[scale=0.31]{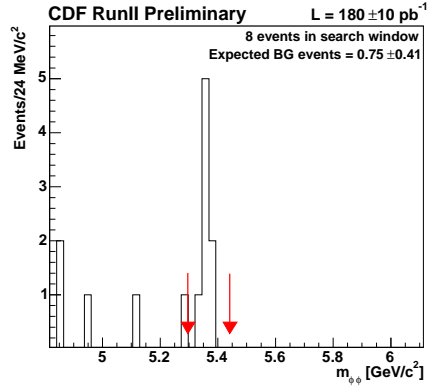} 
\caption{\label{fig_phiphi}
Invariant mass for $\phi \phi$ candidates.} 
\end{center}
\end{figure}

%
Several precision measurements on $B_{u,d}$ meson decays are available, 
yet many crucial theory predictions on $B_s$ mesons, including mixing, 
CP violation and the decay width difference, $\DGs$, are still to be tested. 
\BsVV decays offer insights to both CP violation and \DGs\
thank to the presence of CP-odd and CP-even components in the decay 
amplitude.
\phiphi\ decays are the first charmless \BsVV to be observed.
This channel has been considered extensively in the 
literature~\cite{phiphi-Theory},  even as a probe for New Physics 
along with other $B_s$ modes.
A recent calculation~\cite{phiphi-BR}
predicts a branching ratio of $3.7\cdot 10^{-5}$. 
In the Standard Model (SM) the decay is mediated by $b \myto s\bar{s}s$ 
penguin amplitudes which have shown discrepancies with the SM 
predictions, confirmed by recent
data, for certain CP asymmetry measurements~\cite{Ligety-ICHEP04}.
\phikpm\ decays are mediated by the same quark-level transition and 
have been already studied at B-factories~\cite{pdg2004}.
Measuring precisely rates and CP violation parameters in as many 
such decays as possible may help in determining the source of the above
discrepancies.
\newline
\indent We report the first evidence of the \phiphi\ decay and the 
first measurement, at hadron colliders, of CP-averaged BR and direct CP
asymmetry (\ACP) for \phik\ 
(charge conjugate modes are implied here unless otherwise stated).
In order to cancel the production cross section uncertainty as well as 
to reduce the systematic uncertainty on detector efficiencies, 
the branching ratios are extracted from ratios of decay 
rates of signals and well known $B$ decay modes.
In particular \psiphi\ and \psik decays, characterized by the
same number of secondary vertices and charged tracks in the final state 
as our signals, were 
used respectively in the \phiphi\ and \phik\ analysis. 
\newline
\indent For this measurement we rely on the precision measurement 
of charged particle trajectories reconstructed in the 
central drift chamber (COT) and the silicon detector (SVX\,II).
A complete description of the \cdfii detector can be found 
elsewhere~\cite{cdf}.
The dataset used here was collected by the displaced track trigger,
which is based, at Level~1, on the 
eXtremely Fast Tracker~\cite{xft} and, at Level~2, on the 
Silicon Vertex Tracker~\cite{svt} devices. The trigger
selection is described in detail elsewhere~\cite{D0CP}.
In addition a trigger with relaxed requirements was part of the 
trigger menu with a prescale factor automatically adjusted
to fill the DAQ bandwidth available at low instantaneous luminosity. 
We use $180\, \ipb$ of integrated luminosity, and effectively
only $100\, \ipb$ for the prescaled trigger.
\newline
\indent In this analysis we have used $\phi \myto \KKpm$ and  
$J/\psi \myto \mpmm$ decays. Combinations of three or four tracks with 
$p_T > 0.4\, \GeVc$ are fit to a common vertex.
At least one pair of tracks must satisfy the trigger requirements 
(trigger tracks). To isolate $J/\psi \myto \mpmm$ decays at least
one muon must be identified in the muon detectors.
\begin{figure*}
\hskip -0.2cm
\includegraphics[scale=0.1935]{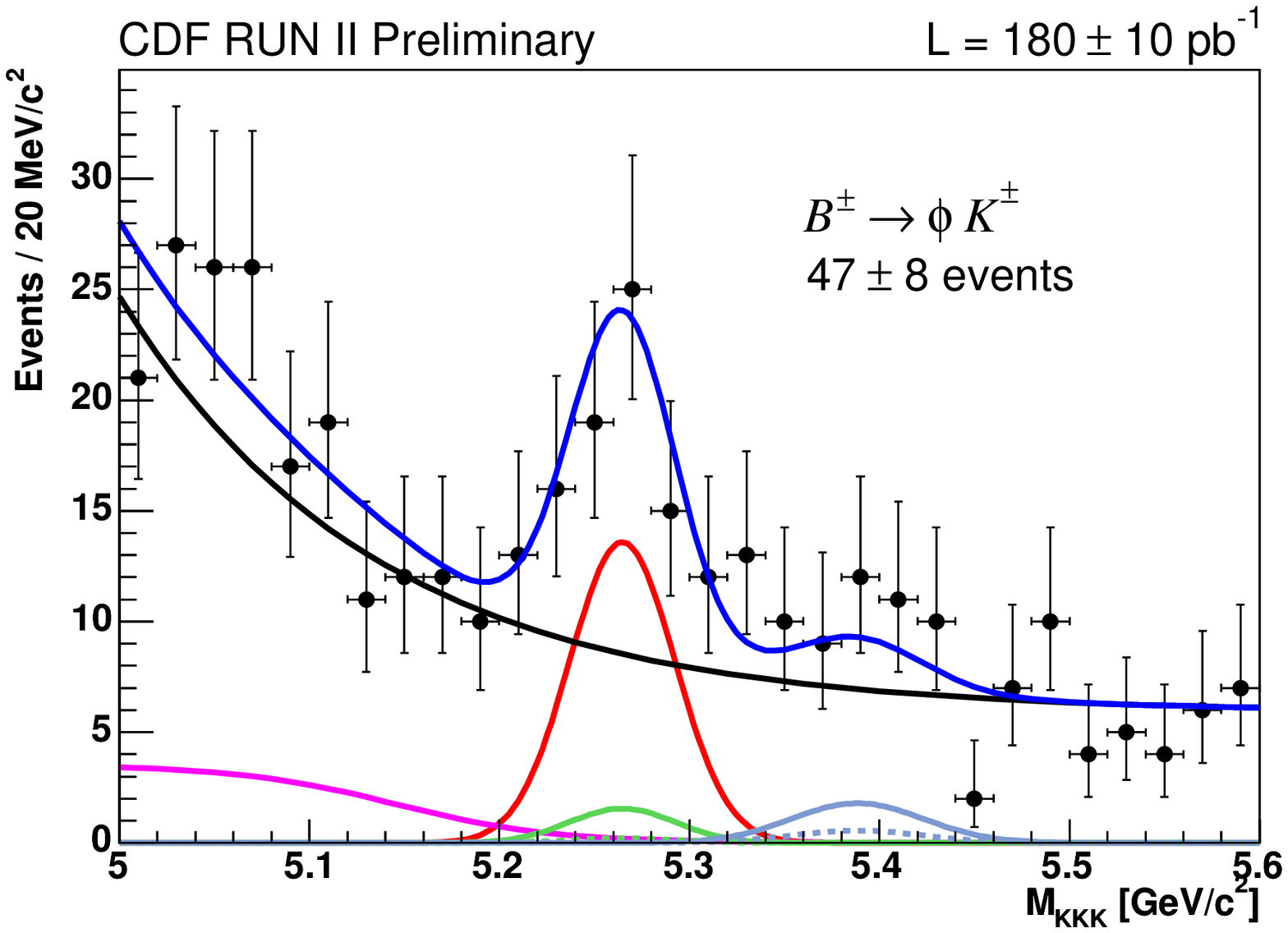} 
\hskip -0.435cm
\includegraphics[scale=0.1935]{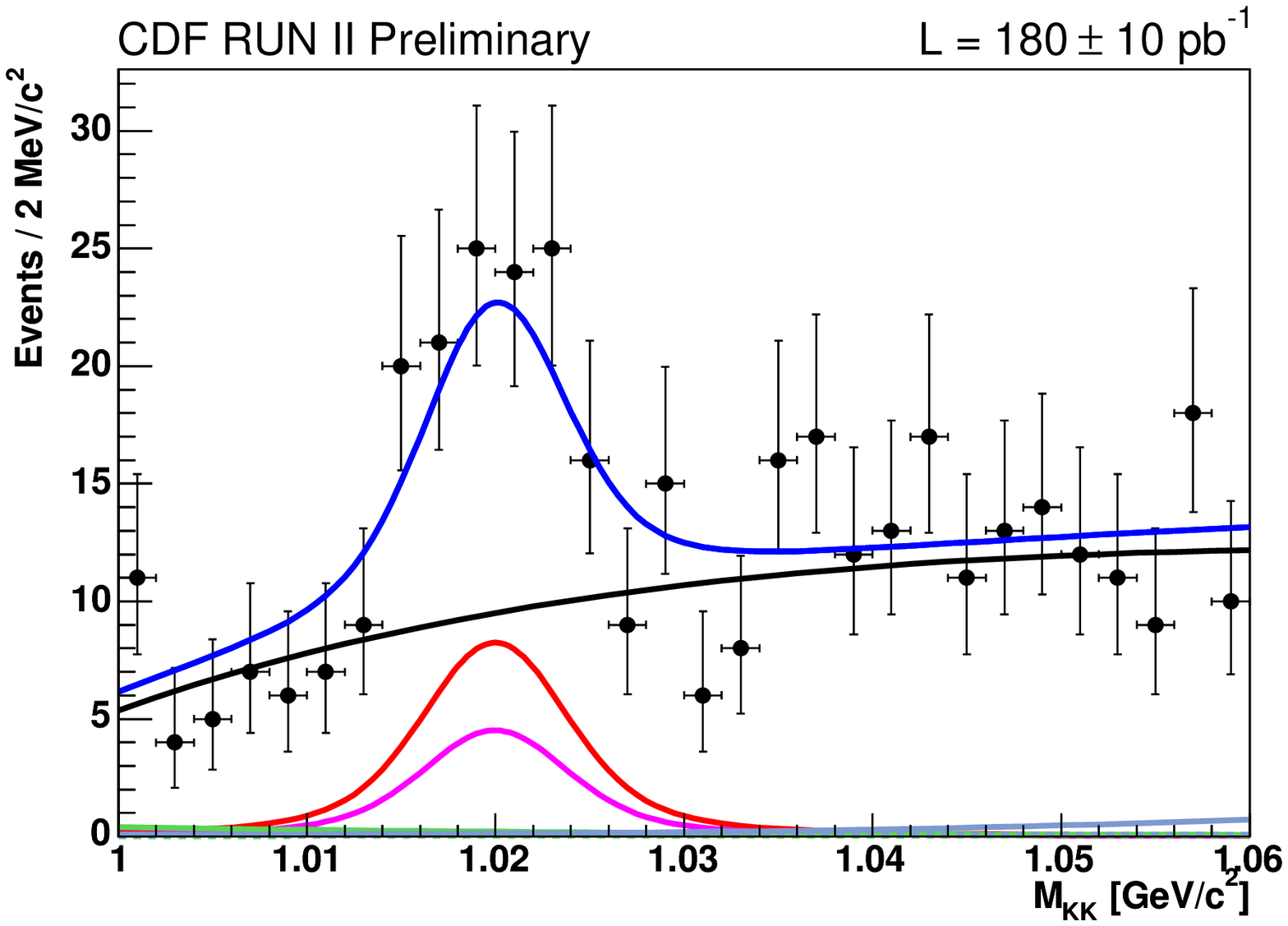} 
\hskip -0.435cm
\includegraphics[scale=0.1935]{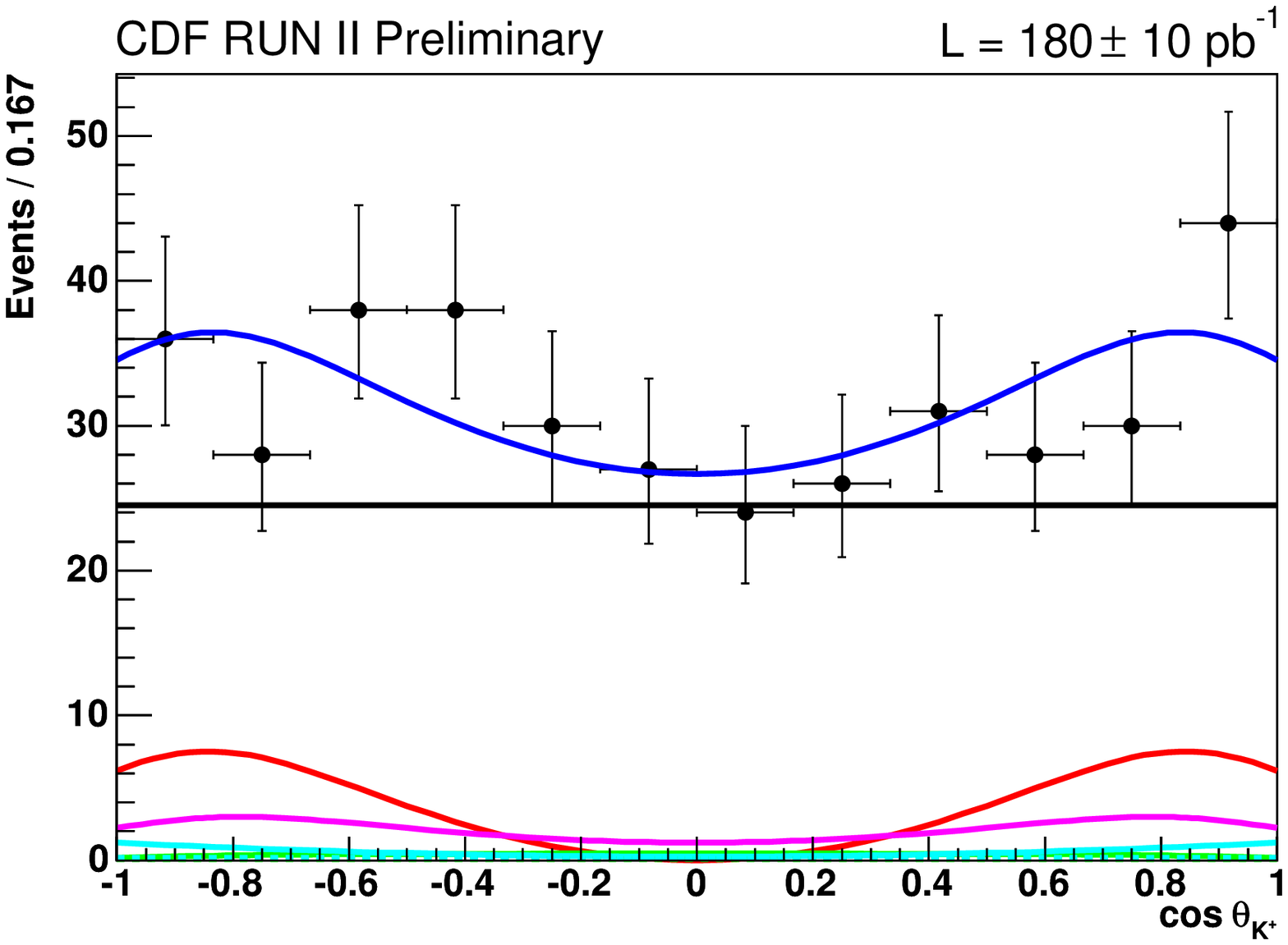} 
\hskip -0.435cm
\includegraphics[scale=0.1935]{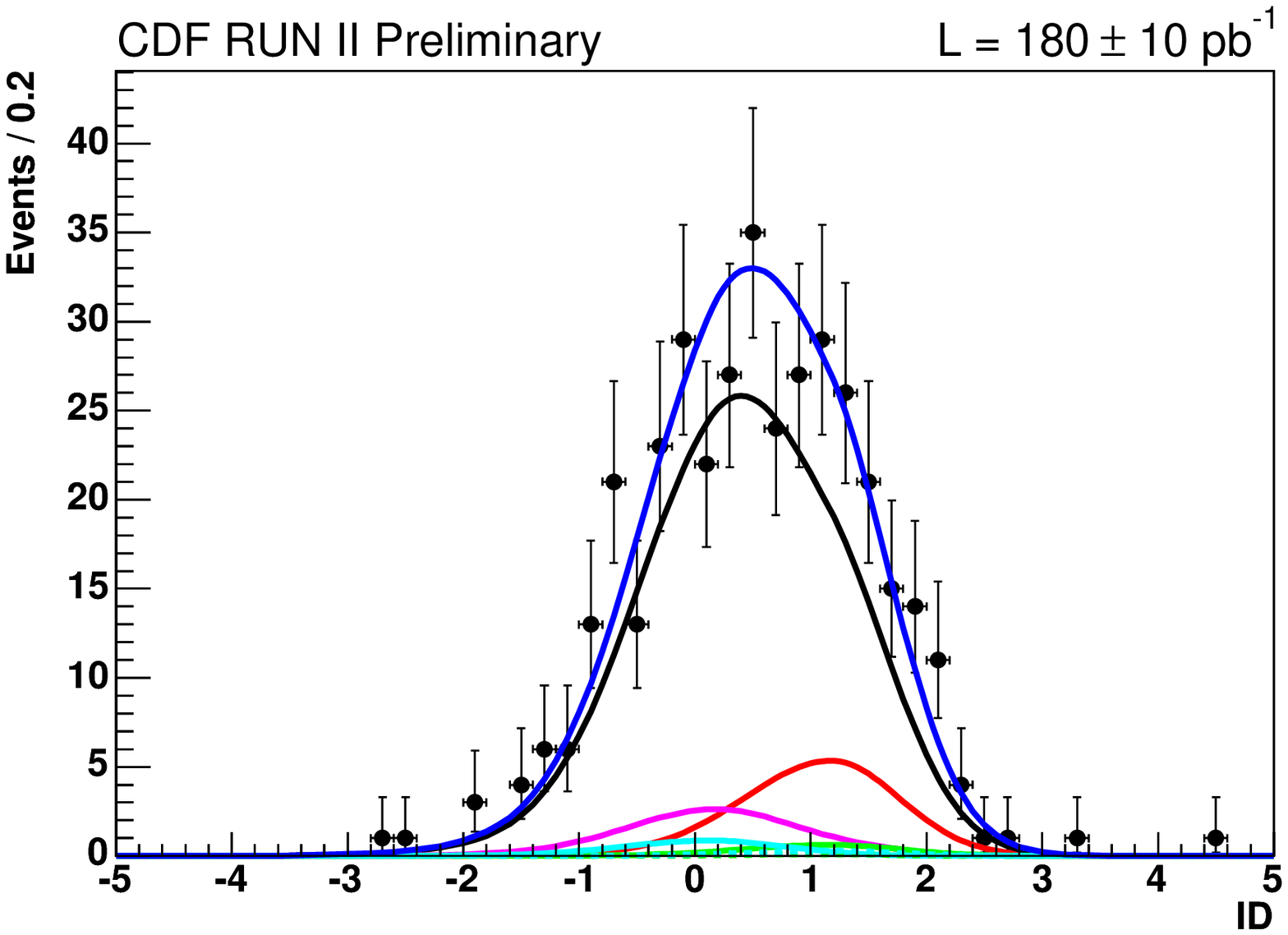} 
\caption{\label{fig_phik}
From left to right $B$ mass ($m_{KKK}$), $\phi$ mass ($m_{KK}$), 
$\phi$ helicity and kaon $dE/dx$ distribution 
of \phik\ candidates along with likelihood projections for: signal(red);  
$b\myto \phi X$ (magenta); combinatorial background (black); 
$B^+\myto f_0(980) K^+$ (green);$B^+\myto K^{\ast 0}\pi^+$ (light blue)
and total (blue).}
\end{figure*}
\newline
\indent
Combinatoric background is reduced by exploiting the long lifetime and 
hard $p_T$ spectrum of $B$ mesons and the isolation of $b$-hadrons inside 
$b$-jets. Requiring the $B$ flight direction 
to point back to the primary 
vertex decreases background from partially reconstructed decays and
selecting good quality vertices background from mis-measured tracks.
The cut values on the discriminating variables were selected maximizing 
$\mathcal {S/\sqrt{S+B}}$ for the already observed \phik\ signal, 
and maximizing $\mathcal{}S/(1.5+\sqrt{\mathcal{B}})$ for \phiphi, 
whose branching ratio was unknown. 
The latter choice is equivalent to maximize the experiment 
potential to reach a $3\sigma$ observation of a new signal~\cite{Punzi}.
In the above expressions the background, $\mathcal{B}$, is represented by 
appropriately normalized data selected in the same way as the signal 
except for requiring the two kaon invariant mass to be in the 
$\phi$ sideband region:$1.04 < m_{KK} < 1.06\, \GeVcc$. 
The signal, $\mathcal{S}$, is derived from a
Monte Carlo (MC) simulation of the \cdfii detector and trigger. For
\phiphi\ a blind analysis was performed. The signal region was hidden
until after all the selection requirements were fixed and background
evaluated.
\newline
\indent
\phik\ candidates are selected requiring: 
vertex goodness-of-fit $\chi^2 < 8$, decay 
length $\Lxy > 350$ $\mu$m, $B^+$ reconstructed impact 
parameter $d_0^B< 100$ $\mu$m, non-trigger track 
transverse momentum $p_T^{\it soft}> 1.3$ $\GeVc$
and impact parameter $d_0^{\it soft}> 120$ $\mu$m. 
We further require that $I_{R<1}> 0.5$, where $I_{R}$ is defined as the
ratio of the $B^+$ candidate $p_{T}$ over the total transverse momenta 
of all tracks within a cone of radius $R=\sqrt{(\Delta\eta^2+\Delta\phi^2)}$ 
around the B flight direction. Moreover, the reconstructed $\phi$ 
mass is required to lie in the interval $1 < m_{KK} < 1.06$ $\GeVcc$ 
\newline
\indent
The signal yield and \ACP, 
defined as
\begin{eqnarray*}
\label{eq:ACPphik}
A_{CP} \equiv \frac{N ( B^-\to\phi K^-) - N ( B^+\to\phi K^+)}
{N ( B^-\to\phi K^-) + N ( B^+\to\phi K^+)} \ ,
\end{eqnarray*}
are extracted simultaneously from an extended 
unbinned maximum likelihood fit on four discriminating variables
(Fig.~\ref{fig_phik}): 
the three-kaon invariant mass, $m_{KKK}$, the invariant mass of the 
$\phi$ candidate, $m_{KK}$, the $\phi$ helicity angle 
and the kaon $dE/dx$ measured in the COT. 
Data are fit to seven categories: signal, partially reconstructed 
$b\myto \phi X$ decays, combinatorial background, 
$B^+\myto K^{\ast 0}(892) \pi^+$ and three $B$ decay modes which peak in 
the signal region, including $B^+\myto f_0(980) K^+$ and 
non-resonant decays. The latter contributions are fixed by
their relative decay rates and detection efficiencies to the
$B^+\myto K^{\ast 0}(892) \pi^+$ one, which is determined from the fit.
A combination of 
Monte Carlo simulation and sideband
data was used to model the signal and background shapes.
The fit returns $N_{\phi K}=47.0\pm 8.4$, $\ACP = -0.07 \pm 0.17$ and
$N_{K^{\ast 0}\pi^+} = 7.8 \pm 6.0$ 
from which we estimate a $B^+\to f_0 K^+$ contamination of 11\%.
\newline
\indent A similar fit uses $m_{\mu\mu K}$ and $m_{\mu\mu}$ on 
\psik\ candidates selected in the same way as in the \phik\ analysis 
above but requiring the invariant mass
of two muons within 100 \MeVcc\ of the $J/\psi$ mass. 
The result is 
$N_{\psi K}=439\pm 22$ and $\ACP = 0.046 \pm 0.050$,
where the error is statistical only.
\newline
\indent
The \phik\ BR ratio is calculated as:
\begin{eqnarray*}
\label{eq:BRphik}
\frac{\BR\left(\phi K^+\right)}{\BR\left(\psi K^+\right)} =
\frac{N_{\phi K}}{N_{\psi K}}
\frac{\BR\left(\psi\to\mu\mu\right)}{\BR \left(\phi\to K K \right)}
\frac{\epsilon_{\psi K}\, \epsilon_{\mu}}{\epsilon_{\phi K}},
\end{eqnarray*}
where $\epsilon_{\psi K}/\epsilon_{\phi K}=0.685\pm 0.015$, 
derived from MC, represents the total detector efficiency ratio 
of the two channels. World average~\cite{pdg2004} 
$\phi$, $J/\psi$ and \psik partial widths are used.
The muon efficiency, $\epsilon_\mu=0.810\pm 0.021$, is determined in a sample 
of inclusive $J/\psi$.
%
%
Systematic uncertainties on signal yield and asymmetry 
are evaluated by varying the parameterizations
used in the likelihood fit, 
including the shape of the $f_0$ resonance. For 
the  branching ratio determination we consider also the uncertainty 
on the relative detection efficiency, which is the dominant one, 
and add it in quadrature to the yield uncertainty. For \ACP\ 
we conservatively assign a 5\% systematic uncertainty from charge 
dependent detector asymmetries using the statistical uncertainty 
on the \psik asymmetry. 
The results are reported in Table~\ref{tab:summResults}.
\begin{table}[h]
\caption{\label{tab:summResults} 
Preliminary CDF results for \phik\ and \phiphi. 
The first uncertainty is statistical, the second systematic.}
\begin{tabular}{lcc}
                          & $ \phik  $  &  $\phiphi $ \\\hline\hline
{ Yield}                  & {\small $47.0 \pm 8.4 \pm 1.4$}               
                          & {\small $7.3 \pm 2.8 \pm 0.4$} \\\hline
{ $\BR \cdot 10^5$}       & {\small $ 0.72 \pm 0.13 \pm 0.07 $}
                          & {\small $ 1.4 \pm 0.6 \pm 0.6$} \\\hline
{       $\ACP$  }         & {\small $-0.07 \pm 0.17 ^{+0.06}_{-0.05}$} 
                          &  --  \\\hline
\end{tabular}
\end{table}
\newline
\indent The \phiphi\ signal is selected requiring two kaon pairs  
with invariant mass within 15 \MeVcc\ of the $\phi$ mass. 
We then apply the following selection:
vertex goodness-of-fit $\chi^2 < 10$, decay length
$\Lxy > 350\,\mu$m, $B_s$ reconstructed impact parameter $d_0^B<80\,\mu$m, 
the minimum momentum of the $\phi$ mesons
$p_T^{\phi} >2.5\,\GeVc$ and minimum impact parameter of the two 
kaons in each of the $\phi$ mesons
($d_0^1> 40\,\mu$m, $d_0^2> 110\,\mu$m).
The \phiphi\ candidate invariant mass distribution is shown in 
Fig.~\ref{fig_phiphi}.
In a region of $\pm 72\, \MeVcc$ around the $B_{s}$ mass,
a window three times the expected mass resolution, 
we count 8 signal candidates. \newline
\indent Two sources of background are expected: combinatorial background 
and \phikst cross-feed with the pion from \kstar decay  
mis-identified as a kaon. The first type of background is studied 
using a background enriched sample where both $\phi$ meson 
candidates have invariant masses lying in the $\phi$ mass sideband region. 
Its contribution in the signal region was estimated as $0.35 \pm 0.37$ events.
From Monte Carlo the expected \phikst background is 
estimated as $0.40 \pm 0.18$ events.
In both cases statistical and systematic uncertainties are included.
\newline
\indent The probability of Poisson fluctuation of background 
to the observed or higher number of events is $1.3 \cdot 10^{-6}$, 
corresponding to a 4.7 $\sigma$ significance. Adding the events 
selected uniquely by the prescaled trigger we find 12 signal candidates
with $1.95\pm0.63$ background, corresponding 
to a 4.8 $\sigma$ significance.\newline
\indent A sample of \psiphi\ is selected requiring one pair of kaons 
and one pair of muons within respectively 15 and 50 \MeVcc\ of the $\phi$ or 
$J/\psi$ mass 
and criteria similar to the \phiphi\ case on decay length and kinematics.
A clean signal of $69 \pm 10(stat.) \pm 5 (syst.)$ \psiphi\ events
is extracted from a fit to the $B_s$ invariant mass distribution. 
The systematic error is evaluated 
using alternative background models. 
From MC simulation we expect a background in the signal peak
of $3.9\pm1.7$ events from \psikst decays with a mis-identified kaon.
\newline
\indent The \phiphi\ decay rate is derived from:
\begin{eqnarray*}
\label{eq:BRphiphi}
\frac{\BR\left(\phi \phi\right)}
     {\BR\left(\psi \phi\right)} =
\frac{N_{\phi\phi}}{N_{\psi\phi}}
\frac{\BR\left(\psi\to\mu\mu\right)}{\BR \left(\phi\to K K \right)}
\frac{\epsilon_{\psi\phi}\, \epsilon_{\mu}}{\epsilon_{\phi\phi}} ,
\end{eqnarray*}
where $\epsilon_{\phi\phi}/\epsilon_{\psi\phi}=0.816\pm 0.015$ and 
$\epsilon_{\mu}\approx 0.92$. We use the world average $\phi$ and 
$J/\psi$ branching ratios~\cite{pdg2004}
and $\BR(\psiphi)=(1.42 \pm 0.51) \cdot 10^{-3}$, obtained 
correcting the CDF measurement~\cite{CDF-psiphiBR} for the current
world average $f_s/f_d$ ratio~\cite{pdg2004}, to finally derive
the result reported in Table~\ref{tab:summResults}.
\begin{figure}
\begin{center}
\includegraphics[width=5.70cm,height=5.40cm]{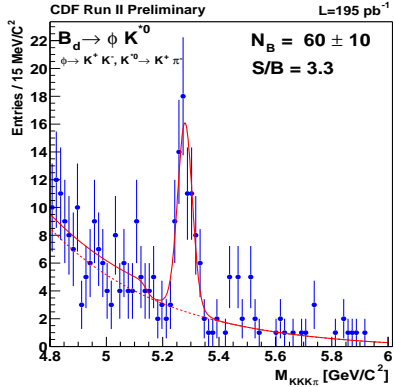} 
\caption{\label{fig_phikstar}
The invariant mass distribution for $\phi \kstar$ candidates using
195 \ipb of data.} 
\end{center}
\end{figure}
\newline
\indent The uncertainty on the \psiphi\ yield  and  
background evaluation contribute 6.9\% to the relative BR systematic 
error. The efficiency ratio is affected by uncertainties in 
the polarization of the decay vector particles and by theory uncertainty 
on \DGs. We conservatively vary the longitudinal polarization of the 
\phiphi\ decay from 0 to 100\% and \DGs\, in the range $0<\DGs/\Gamma_s<0.3$.
Summing in quadrature all contributions 
we estimate a total relative systematic uncertainty error on the 
BR(\phiphi) of 14\%. 
The uncertainty from BR(\psiphi), 36\%, is then the dominant one.
\newline
\indent Thanks to the new displaced track trigger,
CDF is accumulating high quality data on several other charmless decays.  
As an example, Fig.~\ref{fig_phikstar} shows the invariant mass of \phikst
candidates, reconstructed in a similar way as the \phiphi\ above, with a 
signal of $\sim 60$ events and ${\mathcal {S/B}}>3$. It will allow 
precision measurement of polarization and nicely illustrates 
the CDF potential to exploit light vector resonances in $B_s$ decays.
With two to three times the data used here CDF should detect
a signal of \kstarkstarb
with an expected~\cite{phiphi-BR} BR of $\approx 3.7\cdot 10^{-6}$.
\newline
\indent In summary, we have shown the first evidence of \phiphi\ and measure:
$\BR (\phiphi) = (1.4 \pm 0.6(stat.) \pm 0.2 (syst.)\pm 0.5(BR) ) \cdot 10^{-5}$.
For the \phik\ channel we  measure the partial width and \ACP\
which agree with available measurements~\cite{pdg2004} within uncertainties.

\end{document}